\documentclass[%
superscriptaddress,
preprint,
showpacs,preprintnumbers,
 amsmath,amssymb,
aps,
prc,
]{revtex4-1}

\usepackage{graphicx}
\usepackage{dcolumn}
\usepackage{bm}

\begin{document}


\title{Odd-even mass staggering with Skyrme-Hartree-Fock-Bogoliubov theory}

\author{W. J. Chen}
\affiliation{ State Key Laboratory of Nuclear Physics and Technology, School of Physics, Peking University,  Beijing 100871, China}

\author{C. A. Bertulani}
\affiliation{Department of Physics and Astronomy, Texas A{\rm{\&}}M University-Commerce, P. O. Box 3011, Commerce, Texas 75428, USA}

\author{F. R. Xu}
\affiliation{ State Key Laboratory of Nuclear Physics and Technology, School of Physics, Peking University,  Beijing 100871, China}
\affiliation{State Key Laboratory of Theoretical Physics, Institute of Theoretical Physics, Chinese Academy of Sciences, Beijing 100190, China}

\author{Y. N. Zhang}
\affiliation{ State Key Laboratory of Nuclear Physics and Technology, School of Physics, Peking University,  Beijing 100871, China}

\date{\today}

\begin{abstract}
We have studied odd-even nuclear mass staggering with the Skyrme-Hartree-Fock-Bogoliubov theory by employing isoscalar and isovector contact pairing interactions. By reproducing the empirical odd-even mass differences of the Sn isotopic chain, the strengths of pairing interactions are determined. The optimal strengths adjusted in this work can give better description of odd-even mass differences than that fitted by reproducing the experimental neutron pairing gap of $^{120}$Sn.
\end{abstract}

\pacs{21.60.Jz, 21.10.Dr}
\maketitle

One of interesting phenomena in nuclei is the odd-even staggering (OES) of binding energies. It is believed that OES is attributed to the pairing correlation, which plays an important role in nuclear structure \cite{Bohr58, Bohr69}. Numerous microscopic calculations, such as Hartree-Fock + BCS (HF + BCS) or Hartree-Fock-Bogoliubov (HFB) theories, have been performed to investigate the relationship between pairing interaction and OES \cite{Satula98, Xu99, Duguet01, Margueron07, Margueron08, Bertsch09, Yamagami09, Bertulani09, Bertulani12, Changizi15, Cheng15}. The interaction strength, $V_0$, of the pairing interaction is a crucial parameter to understand the nuclear properties of short-range correlations.

Conventionally, the value of pairing strength is obtained by adjusting the average HFB pairing gap of even-even nuclei to fit the experimental odd-even mass differences of the neighboring nuclei. However, the OESs calculated by theoretical masses with the strength determined by the above pairing gap are in some cases substantially different from the odd-even mass differences of the experimental binding energies \cite{Satula98, Xu99, Bertsch09}. The standard pairing strengths adjusted to the average pairing gap in $^{120}$Sn \cite{Doba95} were found to be too small to make a global comparison to the experimental OES data \cite{Bertsch09}. Hence, it is useful to refit the pairing strengths with global calculations to examine the whole system of the OES data.

In the literatures, there are several  measures of  the empirical OES, such as three-point, four-point, and five-point formulae \cite{Bohr69, Satula98, Duguet01}. Here we use the three-point formula $\Delta^{(3)}$  defined as follows,
\begin{equation}
\Delta^{(3)} =  \frac{\pi_{A+1}}{2} [B(N+1,Z) - 2 B(N,Z)   + B(N-1,Z)],
\end{equation}
where $B(N,Z)$ is the binding energy of the $(N,Z)$ nucleus and $\pi_A = (-1)^A$ is the number parity with $A=N+Z$. This second-order difference of binding energies is centered at an odd nucleus, i.e., odd-$N$ nucleus for neutron OES. The $\Delta^{(3)}$ formula can reduce the mean-field contributions to the gap energy \cite{Satula98, Xu99}. For even-$N$ nuclei, the OES is more sensitive to single-particle energies \cite{Satula98}, which is not discussed in this work.

Our investigations are based on the self-consistent Hartree-Fock-Bogoliubov (HFB) calculations with Skyrme energy functionals in the particle-hole channel. We adopt the most commonly used Skyrme parameter sets, SLy4 \cite{Chabanat98}, SkP \cite{Doba84}, SkM* \cite{Bartel82}, UNEDF0 \cite{Kort10} and UNEDF1 \cite{Kort12}. In the particle-particle channel, we employ both isoscalar and isovector density-dependent delta pairing interactions. The isoscalar delta interaction is of the form,
\begin{equation}
V({\bf r}_1, {\bf r}_2) =V_0 \big[ 1 - \eta (\frac{\rho}{\rho_0})^{\gamma}\big] \delta({\bf r}_1 - {\bf r}_2),
\end{equation}
where $V_0$ is the pairing strength, $\eta$ and $\gamma$ are parameters, and $\rho$ is the total density, while $\rho_0$ is the saturation density, which equals to 0.16 fm$^{-3}$. According to the choice of $\eta$, one can obtain different types of pairing, usually called {\it volume, mixed, surface} pairings. The {\it volume} interaction corresponds to $\eta = 0$, which means that there is no explicit density dependence. It mainly acts inside the nuclear volume, while the {\it surface} pairing ($\eta = 1$) is sensitive to the nuclear surface, and the {\it mixed} pairing ($\eta = 0.5$) is a mix in these two pairings. In our calculations we choose $\gamma = 1$.

The isospin-dependent pairing interactions have been proposed to reproduce better pairing gap in nuclei. One kind of isovector pairing, denoted by MSH pairing \cite{Margueron07}, is written as follows,

\begin{eqnarray}
V^{\rm{MSH}}_{pair} ({\bf r}_1, {\bf r}_2)  =  V_0 \big[ 1 - (1-\beta)\eta_s (\frac{\rho}{\rho_0})^{\alpha_s} \nonumber \\
 - \beta \eta_n (\frac{\rho}{\rho_0})^{\alpha_n} \big] \delta({\bf r}_1 - {\bf r}_2),
\end{eqnarray}

where $\rho = \rho_n + \rho_p$, $\beta = (\rho_n - \rho_p)/ \rho$. The parameters are adjusted in the HFB framework to reproduce the pairing gaps in symmetric matter and neutron matter. Here, we adopt the best parametrization with $\eta_s = 0.598$, $\alpha_s = 0.551$, $\eta_n = 0.947$ and $\alpha_n = 0.554$ \cite{Margueron07}.

Another different isovector pairing, denoted by YS pairing \cite{Yamagami09}, is parameterized like this:
\begin{eqnarray}
V^{\rm{YS}}_{pair} ({\bf r}_1, {\bf r}_2) =  V_0 \big[ 1 - (\eta_0 + \eta_1 \tau_3 \beta) (\frac{\rho}{\rho_0}) \nonumber \\
 - \eta_2 (\beta \frac{\rho}{\rho_0})^{2} \big] \delta({\bf r}_1 - {\bf r}_2),
\end{eqnarray}
where $\eta_0 = 0.5$, $\eta_1 = 0.2$, $\eta_2 = 2.5$, and $\tau_3 = -1$ for protons and $1$ for neutrons \cite{Yamagami09}.

We have carried out the HFB calculations with the latest version of HFBTHO \cite{Stoitsov13}. The HFB solver HFBTHO has been developed by implementing the modified Broyden method and shared memory parallelism to accelerate the calculation speed.

The even-even nuclei were first calculated in the HFB framework. We used the orbital space of 20 major harmonic oscillator shells, which is enough for the density functional calculations. For the pairing interactions, we adopted the cutoff energy of 60 MeV.

As for odd-$A$ nuclei, we employed the equal filling approximation (EFA) \cite{Martin08}. Starting from the HFB solution of neighboring even-even nuclei, we select quasi-particle orbitals for the blocking of the odd nucleon. The one-quasiparticle configurations are determined within the blocking energy window $E_{\rm 1qp, win}$, which is smaller than 8 MeV for light nuclei and bigger than 2 MeV for heavy nuclei. In present work, we took $E_{\rm 1qp, win} = 25 / \sqrt{A}$. Finally, we performed unconstrained self-consistent calculations for all candidate configurations and took the minimum energy as the binding energy of the odd-$A$ nucleus.

\begin{table}[h]
\caption{The optimal pairing strengths $V_0$ adjusted to give the best fit to the neutron odd-even staggering of the Sn isotopic chain. The word 'standard' means the pairing strengths fitted by the average pairing gap of $^{120}$Sn. The unit of the pairing strengths is MeV$\cdot$fm$^3$.}

\begin{ruledtabular}
\begin{tabular}{cccccc}

& SLy4 &  SkP &  SkM* & UNEDF0 & UNDEF1 \\
\hline
Standard
 & 283.3 & 213.3 & 233.9 & -- & -- \\
Mixed
 & 310 & 240 & 270 & 225 & 245 \\
MSH
 & 400 & 325 & 360 & 300 & 325 \\
YS
 & 325 & 260 & 290 & 245 & 260 \\

\end{tabular}
\end{ruledtabular}
\end{table}

\begin{figure}
\includegraphics[width=0.5\textwidth, trim=0 200 0 180]{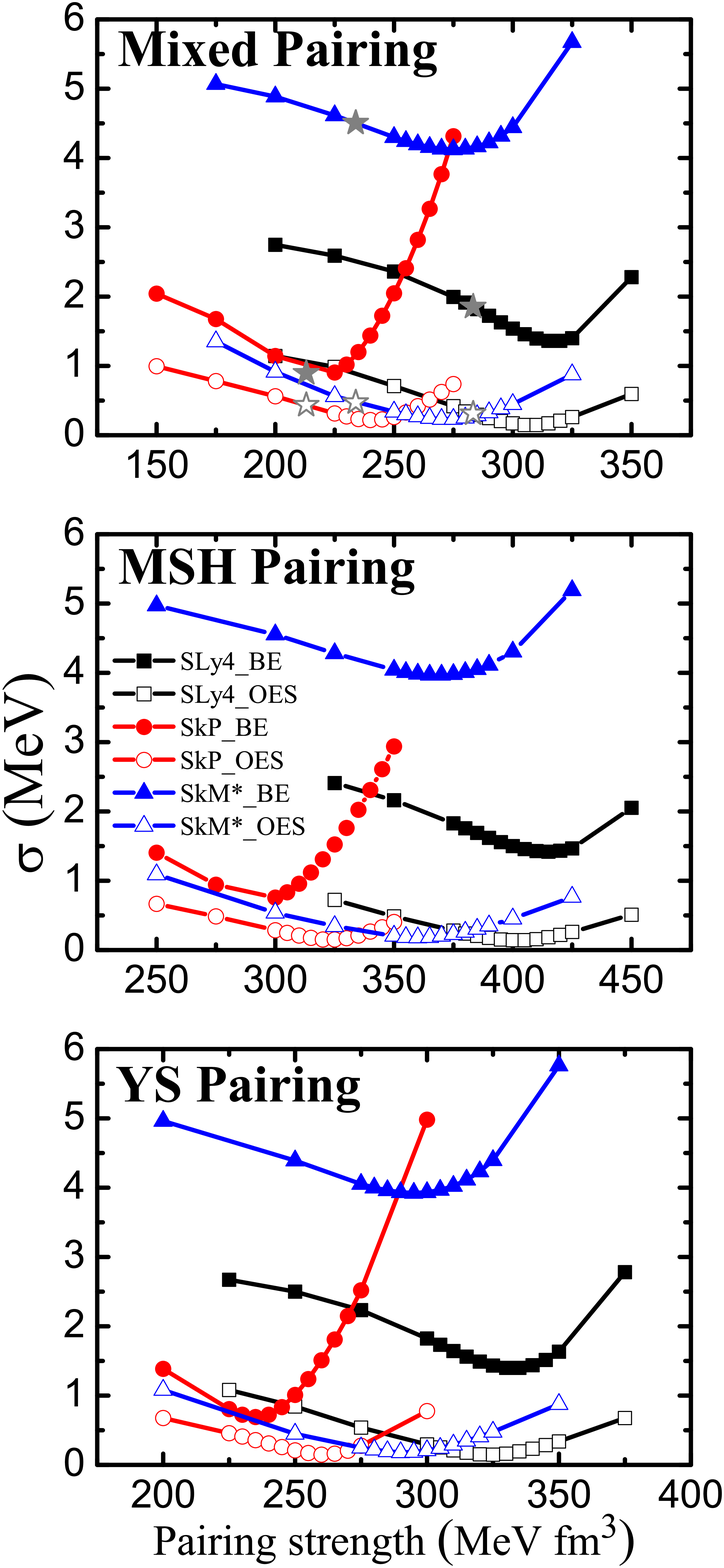}
\caption{(Color online) Root mean square deviation $\sigma$ between HFB calculations and experimental data. The upper, middle and lower panels correspond to HFB calculations with Mixed pairing, MSH pairing and YS pairing, respectively. Filled symbols stand for the mean square deviation of binding energies of the Sn isotopic chain, while open ones for the deviation of OES. The gray stars in the upper panel represent the results calculated with the strength fitted by the average pairing gap of $^{120}$Sn. See text for details.}
\end{figure}

As mentioned before, the pairing strengths should be refitted to examine the whole system of the OES data. In principle, global calculations are needed. However, such calculations for different Skyrme forces and pairing interactions are very time consuming. We only adjust the pairing strengths to give the best fit to the OES of the semimagic Sn isotopes with neutron number ranging from 49 to 85. In fact, Sn isotopes are known to be excellent laboratories for comparison with mean field calculations and OES effects \cite{Bertulani09}. In present work, the proton pairing strengths are identical to the neutron ones.

Table I lists the optimal pairing strengths for the five different Skyrme forces with different pairing interactions. The results for the root mean square deviation of our calculations are shown in Figure 1. The root mean square deviation $\sigma$ of OES between HFB calculations and experimental data is defined as
\begin{equation}
\sigma({\rm OES}) = \sqrt{ \sum_{i=1}^{N_i} \vert \Delta_i^{(3)}({\rm HFB}) - \Delta_i^{(3)}({\rm Exp}) \vert ^2 / N_i}
\end{equation}
where $N_i$ is the number of data points. The experimental masses are taken from Ref. \cite{Nubase12}.

\begin{figure}[h]
\includegraphics[width=0.45\textwidth, trim=0 100 0 50]{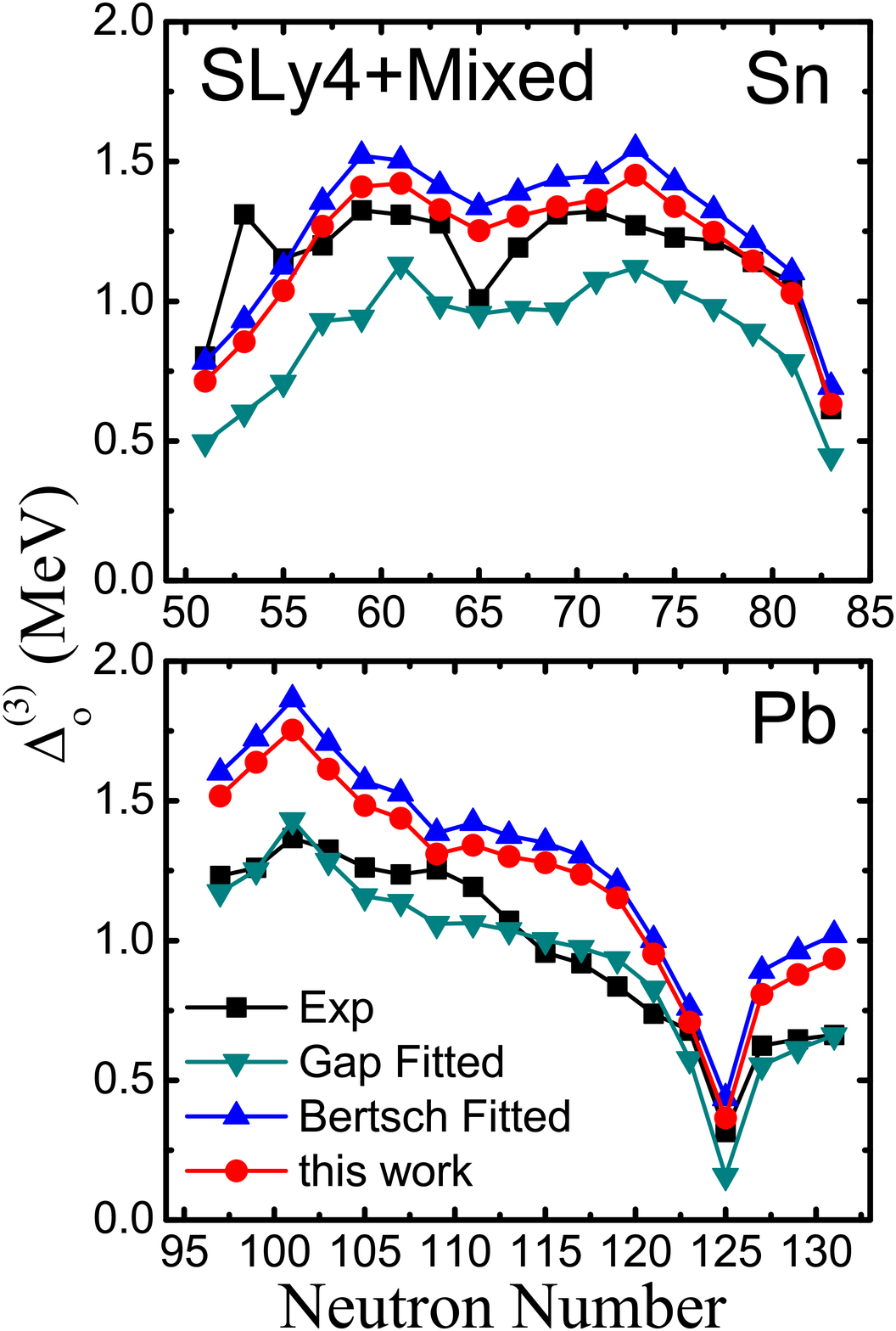}
\caption{(Color online) Calculated and experimental values of $\Delta_o^{(3)}$ for the semimagic Sn and Pb isotopic chains. The SLy4 interaction is adopted with mixed pairing in the HFB framework for different pairing strengths. The dark cyan down triangles shows the results with the standard pairing strength listed in Table 1 and the blue up triangles stand for the results with the pairing strength fitted by Bertsch et. al. \cite{Bertsch09} }
\end{figure}

In Figure 1, we compare the results for the three commonly used Skyrme parameter sets, SLy4, SkP and SkM*. It seems that the optimal pairing strength to obtain the best fit to binding energies are different from that for OES. The minima of the fitting curve of binding energies for SLy4 and SkM* forces with mixed pairing are slightly bigger than that of OES, while for the SkP force, it shows an opposite result. The small differences of the minima for binding energies and OES come from the definition of the root mean square deviation $\sigma$. From Eqs.(1) and (5), we can see that the deviation $\sigma$ of OES contains the correlations between neighborhood nuclei, which is not included in the deviation of binding energies.

\begin{figure}[h]
\includegraphics[width=0.45\textwidth, trim=0 100 0 50]{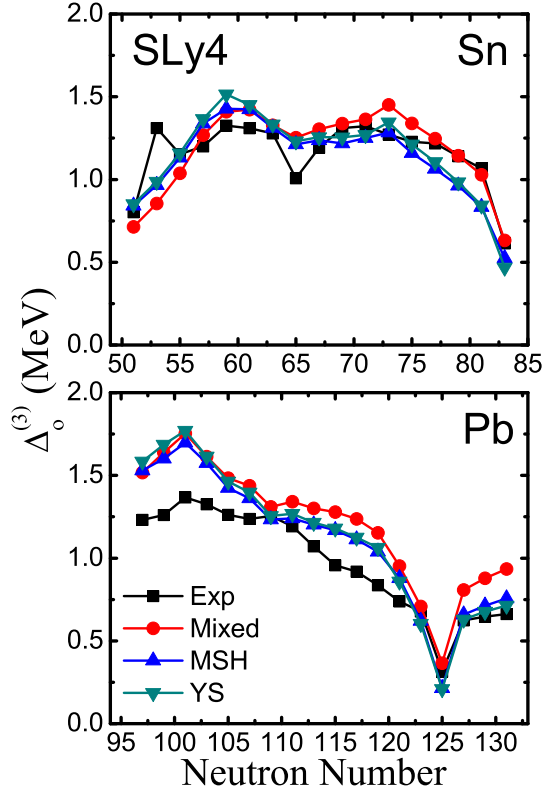}
\caption{(Color online) Calculated and experimental values of $\Delta_o^{(3)}$ for neutrons in the Sn and Pb semimagic isotopic chains. SLy4 is adopted together with the mixed pairing (Eq.(2)) and two isospin dependent pairing (MSH in Eq.(3) and YS in Eq.(4)) interactions in the HFB model.}
\end{figure}

For the SkP force, the different results from SLy4 and SkM* forces may come from its parametrization procedure. The SkP parameters are optimized together with pairing force, which is similar to the recently published parameter sets UNEDF0 and UNEDF1 \cite{Doba84, Kort10, Kort12}. In fact, we also adjusted the pairing strengths for UNEDF0 and UNEDF1 forces with mixed pairing interaction. The results show that as the pairing strength increases, the root mean square deviation of binding energies increases sharply. Take UNEDF0+Mixed as an example, the optimal pairing strength for OES is at 225 $\rm{MeV} \cdot \rm{fm}^{3} $, where the mean square deviation $\sigma$ of binding energies is 3.18 MeV. We remind the reader that the fitting of pairing strengths should be done with caution for some Skyrme forces.

Figure 2 shows the OESs of two semimagic isotopic chains Sn and Pb with three pairing strengths fitted by different ways. The overall trend is reproduced for all three treatments of pairing. For Sn isptopes, the flatness up to the quenched gap at N = 83 is well consistent to experimental data, and for Pb isotopes, the downsloping trend  up to gap at N = 125 is also reproduced. The pairing strength fitted by overall systematics by Bertsch et. al.\cite{Bertsch09} gives higher average OES in both spherical chains, while the strength adjusted by reproducing the pairing gap of $^{120}$Sn gives too small values in the Sn isotopic chain. Our results are between these two values.

We have also compared the results calculated by the optimal strengths with three different pairing interactions, shown in Figure 3. The MSH and YS pairing interactions are essentially the mixed-type contact pairing interactions with isospin dependence. The overall trends for the three different pairing interactions are similar. For nuclei with neutron excess, the OESs with MSH and YS pairing interactions are smaller than with mixed pairing interaction. The isospin dependent pairing interactions flatten the odd-even mass differences as a function of neutron number, which is consistent with the results in Ref. \cite{Margueron08}.

In summary, we have investigated the neutron OESs of Sn and Pb isotopes using self-consistent Skyrme-Hartree-Fock-Bogoliubov theory with SLy4, SkP, SkM*, UNEDF0 and UNEDF1 forces together with mixed pairing and two different isospin dependent pairing interactions. The pairing strengths are adjusted by reproducing the empirical OESs of Sn isotopes. The pairing strengths necessary to obtain the best fit to the binding energies are  different from that for OES. We reproduced the flatness of the OES due to the isospin effects and compared to the results with isoscalar pairing interactions. \\

This work was partially supported by the CUSTIPEN (China-U.S. Theory Institute for Physics with Exotic Nuclei) under DOE grant number DE-FG02-13ER42025; the  U.S. NSF Grant No. 1415656, the U.S. DOE grant No. DE-FG02-08ER41533 ; the National Key Basic Research Program of China under Grant No. 2013CB83440; the National Natural Science Foundation of China under Grants No. 11235001, No. 11320101004; the Open Project Program of State Key Laboratory of Theoretical Physics, Institute of Theoretical Physics, Chinese Academy of Sciences, China (No. Y4KF041CJ1).

\bibliography{prc_pairing_isospin}

\end{document}